\documentclass[structabstract,twocolumn]{aa}  
\usepackage{natbib}
\usepackage{graphicx}
\usepackage{txfonts}

\usepackage{amssymb}
\usepackage[notref,notcite]{} 
\usepackage{color}

\newcommand\U[1]{{\,\rm #1}}
\newcommand\kms{\rm ~km\,s^{-1}}
\newcommand\cmc{cm^{-3}}

\newcommand\sizeFig{1.0}

\newcommand\rs[1]{_\mathrm{#1}}

\newcommand\Vsh{V\rs{sh}}

\newcommand\rhoztot{\rho\rs{0,tot}}
\newcommand\rhozion{\rho\rs{0,ion}}

\newcommand\Te{T\rs{e}}

\newcommand\epsCR{\epsilon\rs{CR}}
\newcommand\PCR{P\rs{CR}}

\newcommand\hN{h\rs{N}}
\newcommand\betadown{\beta\rs{down}}

\newcommand\SEin{SE$\rs{in}$}
\newcommand\SEout{SE$\rs{out}$}

\begin{document}

\title{Cosmic ray acceleration and Balmer emission from RCW~86 (G315.4\,--\,2.3)}

\author{G. Morlino\inst{1,2}\fnmsep\thanks{email: morlino@arcetri.astro.it}, 
            P. Blasi\inst{1,3}, R. Bandiera\inst{1} \and E. Amato\inst{1}
            }
\institute{$^1$INAF/Osservatorio Astrofisico di Arcetri, Largo E. Fermi, 5, 50125, Firenze, Italy \\
              $^2$APC, AstroParticule et Cosmologie, Universit\'e Paris Diderot, CNRS/IN2P3, CEA/Irfu, Observatoire de Paris, Sorbonne Paris Cit\'e, 10, rue Alice Domon et L\'eonie Duquet, F-75205 Paris Cedex 13, France \\
	      $^3$INFN/Gran Sasso Science Institute, viale F. Crispi 7, 67100 L'Aquila, Italy
               }

\date{Received 5 November, 2013; accepted 11 December, 2013}

 
  \abstract
{Observation of Balmer lines from the region around the forward shock of supernova remnants (SNR) may provide valuable information on the shock dynamics and the efficiency of particle acceleration at the shock.}
{We calculated the Balmer line emission and the shape of the broad Balmer line for parameter values suitable for SNR RCW~86 (G315.4\,--\,2.3) as a function of the cosmic-ray (CR) acceleration efficiency and of the level of thermal equilibration between electrons and protons behind the shock. This calculation aims at using the width of the broad Balmer-line emission to infer the CR acceleration efficiency in this remnant.}
{We used the recently developed nonlinear theory of diffusive shock-acceleration in the presence of neutrals. The semianalytical approach we  developed includes a description of magnetic field amplification as due to resonant streaming instability, the dynamical reaction of accelerated particles and the turbulent magnetic field on the shock, and all channels of interaction between neutral hydrogen atoms and background ions that are relevant for the shock dynamics.} 
{We derive the CR acceleration efficiency in the SNR RCW~86 from the Balmer emission . Since our calculation used recent measurements of the shock proper motion, the results depend on the assumed distance to Earth. For a distance of 2~kpc the measured width of the broad Balmer line is compatible with the absence of CR acceleration. For a distance of 2.5~kpc, which is a widely used value in current literature, a CR acceleration efficiency of 5\,--\,30\% is obtained, depending upon the electron-ion equilibration and the ionization fraction upstream of the shock. By combining  information on Balmer emission with the measured value of the downstream electron temperature, we constrain the CR acceleration efficiency to be $\sim 20\%$. 
}
   {}
\keywords{acceleration of particles -- cosmic rays -- Balmer emission --
          SNR: RCW~86 }

\maketitle

\section{Introduction}

The supernova remnant (SNR) RCW~86 (G315.4\,--\,2.3) has been the subject of much attention, especially after its detection as a TeV gamma-ray source by means of the HESS Cherenkov telescope \citep{HESS09}. The gamma-ray emission together with the detection of X-ray non-thermal synchrotron radiation from the northeastern part of the remnant \citep{Bamba00, Borkowski01, Vink06} suggest that this remnant may be an efficient cosmic ray (CR) accelerator. If the gamma-ray emission were confirmed to arise from pion production and decay, this evidence would support the SNR paradigm for the origin of the bulk of Galactic CRs. 

In the context of this paradigm, particle acceleration in SNRs takes place at shocks associated with the supernova explosion, and is described by the non linear theory of diffusive shock acceleration \cite[see][for a review]{maldrury}. Energy and momentum conservation at the shock in the presence of accelerated particles leads to two straightforward conclusions: 1) since part of the energy is channeled into particle acceleration, the thermal energy (hence the temperature) of the downstream gas is expected to be lower than in the absence of CRs; 2) the dynamical reaction of accelerated particles induces the formation of a precursor upstream of the shock, which results in deviation of the spectrum of accelerated particles from a power-law behavior. 

The acceleration efficiency typically required for SNRs to be the sources of CRs is $\sim 10\%$. In the pioneering paper by \cite{Chevalier78}, it was first suggested that if the supernova explosion occurs in a partially ionized medium, Balmer line profiles should consist of a broad component and a narrow component. The line width of the broad component, observed in proximity of the collisionless shock, could provide information on the temperature of the downstream plasma. Since this temperature is lower if particle acceleration is efficient, this would also provide information on the efficiency of CR acceleration. If there were efficient acceleration, the Balmer line emission would still present a broad and a narrow component, but the former would be narrower and the latter broader than they would be in the absence of particle acceleration. An important role in the description of a collisionless shock propagating in a partially ionized medium, and hence on the points discussed above, is played by the level of thermal equilibrium between electrons and ions \cite[see][for a recent review]{2013SSRv.tmp.75G}.

Despite the simple expectations illustrated above, a comprehensive description of the behavior of a collisionless supernova shock in the presence of neutrals, for arbitrary shock speeds, has only recently been put forward. A first effort to include neutrals in the shock acceleration theory was made by \cite{Wagner09} using a two-fluid model to treat ions and CRs, but neglecting the dynamical role of neutrals. A different model was proposed by \cite{Raymond11}: here momentum and energy transfer between ions and neutrals are included, but both the CR spectrum and the profile of the CR-precursor were assumed {\it a priori} rather than calculated selfconsistently. 
\cite{paperI} proposed a semianalytical kinetic calculation that returns the distribution function of neutral hydrogen atoms in the absence of accelerated particles and all thermodynamical properties of the ion plasma. These authors showed that a substantial fraction of neutral atoms that suffer a charge-exchange (CE) reaction downstream produce neutral atoms that move upstream and can release energy and momentum in the upstream plasma, thereby heating it. \cite{paperII} showed that this phenomenon (hereafter {\it neutral return flux}) creates an intermediate component of the Balmer line. Both these calculations were used by \cite{paperIII} to generalize the non linear theory of diffusive shock acceleration \cite[]{AmatoBlasi05,AmatoBlasi06,Caprioli08} to include the effects associated with the presence of neutrals. This theory allows one to calculate the spectra of accelerated particles, the magnetic field amplification at the shock, and the modifications induced on the shock by neutral atoms. The processes of CE and ionization, as well as excitation of atoms as due to collisions with both ions and electrons, are included in the calculations: all the elements are provided to accurately calculate the Balmer emission, as discussed by \cite{paperII}. 

In this paper we apply the theory we recently developed to the SNR RCW~86. We concentrate on the eastern part of the remnant because this is the region showing the highest shock speed according to measurements of optical proper motion, hence it appears as the best candidate for efficient particle acceleration.
The width of the broad Balmer line from the northeastern (NE) part of this remnant was measured by \cite{Helder09}. The authors combined proper motion measurements of non-thermal X-ray filaments observed with Chandra with optical measurements made with the VLT of the width of the broad $H\alpha$ component from the same region, showing that the temperature behind the shock is lower than expected based on Rankine-Hugoniot relations. They concluded that a sizable fraction of the energy ($\gtrsim 50\%$) is being channeled into CRs. In fact, this conclusion has been somewhat modified by the same authors in a recent paper \cite[]{Helder13}, where they used new measurements of the proper motion of optical filaments, which imply a lower shock velocity than inferred from the X-ray proper motion. Using the new measurements, the CR acceleration efficiency that the authors estimate is lower and depends on the assumed distance $d$ of the remnant, which is compatible with zero for $d\lesssim 2.5\U{kpc}$.

Here we used the theory illustrated in \cite{paperIII} to calculate the CR acceleration efficiency that best describes the measured broad Balmer line width. Our calculations show that the inferred CR acceleration efficiency depends on the distance to RCW~86. For a distance of 2~kpc the observed width of the broad Balmer line is still compatible with the absence of CR acceleration, while for a distance of 3~kpc all measurements require effective CR acceleration. We adopted a fiducial value of the distance to RCW~86 of 2.5~kpc and carried out our calculations for this case. We focused our attention on a specific sector of the SE rim observed by \cite{Helder13,Helder11}, which was labeled by these authors as $\rm SE_{out}$ and appears most promising in terms of CR acceleration. The estimated CR acceleration efficiency depends on the level of electron-ion equilibration and varies between 5\% and 30\% for a neutral fraction ($\hN$) of 10\%. 
When combined with the measurement of the electron temperature as obtained from X-ray observations by \cite{Helder11}, the estimated CR acceleration efficiency is $\sim 20\%$ in the cases of interest. 

The paper structure is as follows: in \S~\ref{sec:dynamics} we discuss the observational constraints on the physical parameters of the SNR RCW~86, with particular care for the distance of this remnant from the Sun, the shock velocity, and the gas density. The results of our calculations are illustrated in \S~\ref{sec:results}, where we calculate the CR acceleration efficiency and the shape of the Balmer-line emission. We summarize our results in \S~\ref{sec:conclusion}.

\section{Constraints on physical parameters}
\label{sec:dynamics}

To calculate of the expected Balmer emission, we need to discuss the range of allowed values for parameters such as shock velocity, distance from the Sun, gas density, neutral fraction upstream of the shock, and the ratio of electron to ion temperatures. The origin of RCW~86 (G315.4\,--\,2.3) has been much debated in the literature: there are serious difficulties in reconciling the young age of RCW~86 with its relatively large size \cite[]{Williams11}. If the distance to this remnant (which is $40'$ in diameter) is $\sim 2.5\U{kpc}$, the average speed of the shock during its evolution can be estimated to be $\sim 7800\U{\kms}$. However, the shock speed in most of the remnant is about one order of magnitude lower than this value. \cite{Vink97} suggested that the discrepancy could be explained by assuming that the supernova explosion took place in a low-density cavity, possibly associated to the progenitor of the supernova itself, so that the shock would have propagated in a low-density medium for a long time and eventually have slowed down after encountering the denser interstellar medium. 
A similar hypothesis was put forward by \cite{Gvaramadze03}, who proposed that RCW~86 is the result of an off-centered cavity SN explosion of a moving star (supported by the discovery of a candidate neutron star in the ``proper'' place).
More recently, \cite{Badenes07} suggested that RCW~86 may be the result of a single degenerate Type~Ia SN that exploded into a low-density bubble ($n_{\rm bubble}\sim 10^{-3}\U{\cmc}$) created by the wind of the companion star. In conclusion, the most likely scenario seems to be one in which this SN exploded into a low-density cavity, and the low shock speed observed today (especially in the SW region) can be explained assuming that the shock has by now encountered the dense shell at the edge of the cavity, with a density $\sim 1\U{\cmc}$. 

The estimates of gas density in the region of RCW~86 were summarized by \cite{Dickel01}: the available estimates, based on data in different wavebands, taken in different regions of the remnant \citep{Vink97, Claas89, Greidanus90, Smith97} seem to support a scenario in which RCW86 has been expanding in an underdense bubble and is now impacting a higher-density clump in the SW. The difference between the NE and SW is also consistent with a factor of 4 difference in ambient density between the two sides \citep{Petruk99}. More recently, \cite{Williams11}, fitting the infrared flux ratios with models of collisionally heated ambient dust, found post-shock gas densities of 2.4 and $2.0\U{\cmc}$ in the SW and northwest portions of the remnant, respectively. Overall, a reasonable estimate for the density of the unshocked gas appears to be around $1\U{\cmc}$, and we adopted this value in all the calculations presented here. We note that the main conclusions presented here are not strongly affected by the value of the total upstream density.

The distance to SNR RCW~86 is very uncertain, but generally bound between 2 and 3~kpc. \cite{Westerlung69} inferred a distance of 2.5~kpc from the possible association with the location of a nearby OB association. A distance of 2.3\,--\,2.7~kpc was inferred from an analysis of radio data \cite[see][]{Westerlung69}. \cite{Rosado96} found a distance of $2.8\pm0.4\U{kpc}$ from kinematic arguments, while \cite{Leibowitz83} used optical extinction to estimate a distance of 3.2~kpc. A safe value appears to be $2.5\pm0.5\U{kpc}$, but in our calculations we discuss the implications of this uncertainty on the results. 

No reliable data on the ionization fraction was found in the literature, therefore we consider here the values 50\% and 10\% as reasonable estimates.

\section{FWHM of the broad Balmer line and CR acceleration}
\label{sec:results}

The calculations of the shock structure of SNR RCW~86 were carried out by using the formalism of \cite{paperIII}, where both the CR reaction and the shock modification induced by neutral hydrogen are taken into account. As discussed by \cite{paperIV}, the kinetic approach adopted for the description of the neutral distribution function is crucial in this type of calculations. Since there is no guarantee that thermalization between ions and neutrals is ever reached, assuming that the neutrals become maxwellian with the same temperature as the ions can lead to overprediction of the width of the Balmer line. The situation is especially critical for shocks faster than 2500 km/s, where neutrals are ionized before they have time to thermalize, but it is generally true that only by means of a kinetic description it is possible to obtain a reliable estimate of the acceleration efficiency based on Balmer-line emission. 

\begin{figure}
\begin{center}
{\includegraphics[width=\sizeFig\linewidth]{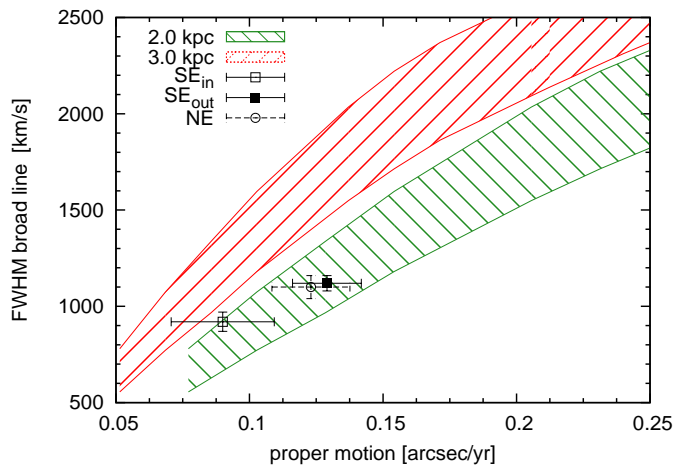}}
  \caption{FWHM of the broad Balmer line (without CRs) as a function of the proper motion for a distance of 2 (green) and 3~kpc (red) from the Sun. The level of electron-ion equilibration varies in each of the hatched regions, with $\beta_{\rm down}=0.01$ and $\beta_{\rm down}=1$ at the upper and lower bound, respectively. Data points show the measurements of proper motion and width of Balmer emission in different regions of the shock, as reported by {\protect \cite{Helder13}}.} 
 \label{fig:FWHM_d2-3}
\end{center}
\end{figure}

In the absence of accelerated particles, the calculations are as described by \cite{paperI}. The FWHM of the broad Balmer line, assuming that no particle acceleration is taking place, is shown in Fig. \ref{fig:FWHM_d2-3} as a function of proper motion for the two cases of a distance to the SNR of 2 (green) and 3~kpc (red) from the solar system. The hatched regions illustrate the dependence of the calculated FWHM on the electron-ion equilibration. The data points show the measured FWHM in the SE and NE rims as found by \cite{Helder13}. For the SE region, \cite{Helder11} provided measurements in two different sectors of the shock, which they labeled \SEin\ and \SEout\ .  A given proper motion corresponds to a higher shock velocity for a larger distance, hence it is not surprising that the expected FWHM of the broad Balmer line is larger when the source is assumed to be more distant. For a distance of 2~kpc, our calculations show that the measured FWHM is compatible with the absence of CR acceleration in RCW~86 for all regions where the Balmer emission has been measured, especially if electrons do not equilibrate efficiently with ions. On the other hand, for a source distance of 3~kpc, all data points suggest some level of CR acceleration, since the measured FWHM is systematically smaller than the FWHM as derived from $\Vsh$, even in the limit case of full electron-ion equilibration. In the rest of this section we  assume a distance of 2.5~kpc.

\begin{figure}
\begin{center}
{\includegraphics[width=\sizeFig\linewidth]{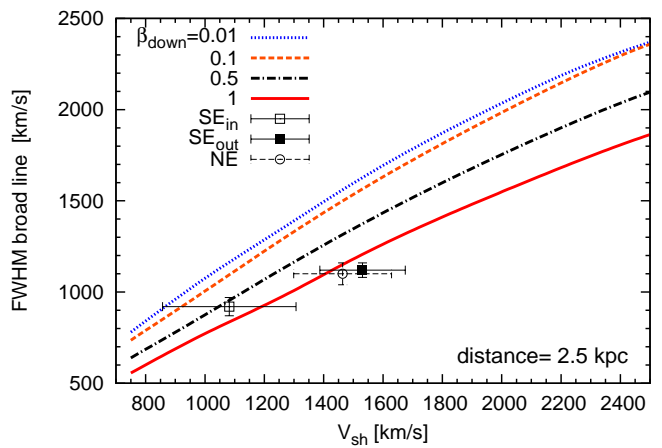}}
  \caption{FWHM of the broad Balmer line in the absence of CR acceleration and for different electron-ion equilibration levels. Here the distance is assumed to be 2.5~kpc. The assumed neutral fraction is $\hN=50\%$ (but the theoretical curves depend very little on $\hN$ when no CRs are present). Data points are the same as in Fig.~\ref{fig:FWHM_d2-3}. }
  \label{fig:FWHM_d2.5}
\end{center}
\end{figure}

In Fig. \ref{fig:FWHM_d2.5} we show the results of our calculations of the FWHM in the absence of particle acceleration as a function of the shock velocity as deduced from the proper motion for a source distance of 2.5~kpc and neutral fraction $\hN=50\%$ (in the absence of CR acceleration the results depend very weakly on $\hN$). From top to bottom the curves refer to a ratio of electron-to-ion temperature $\betadown \equiv T_e/T_p=0.01$, 0.1, 0.5, and 1 (as labeled in the figure). The data points are again taken from the work by \cite{Helder13}. This plot reinforces the conclusion that the requirement for efficient CR acceleration is especially strong in the case of poor equilibration between electrons and ions, while the results are marginally consistent with no CR acceleration for the case of full equilibration (solid red line). The error bar on the proper motion in the \SEin\ region (see Fig.~1 in \cite{Helder11}, corresponding to Filament \#3 in Table 2 of \cite{Helder13}) is so large that the measured FWHM is consistent with no particle acceleration for all levels of electron-ion equilibration. 

\begin{figure*}
\begin{center}
{\includegraphics[width=\sizeFig\linewidth]{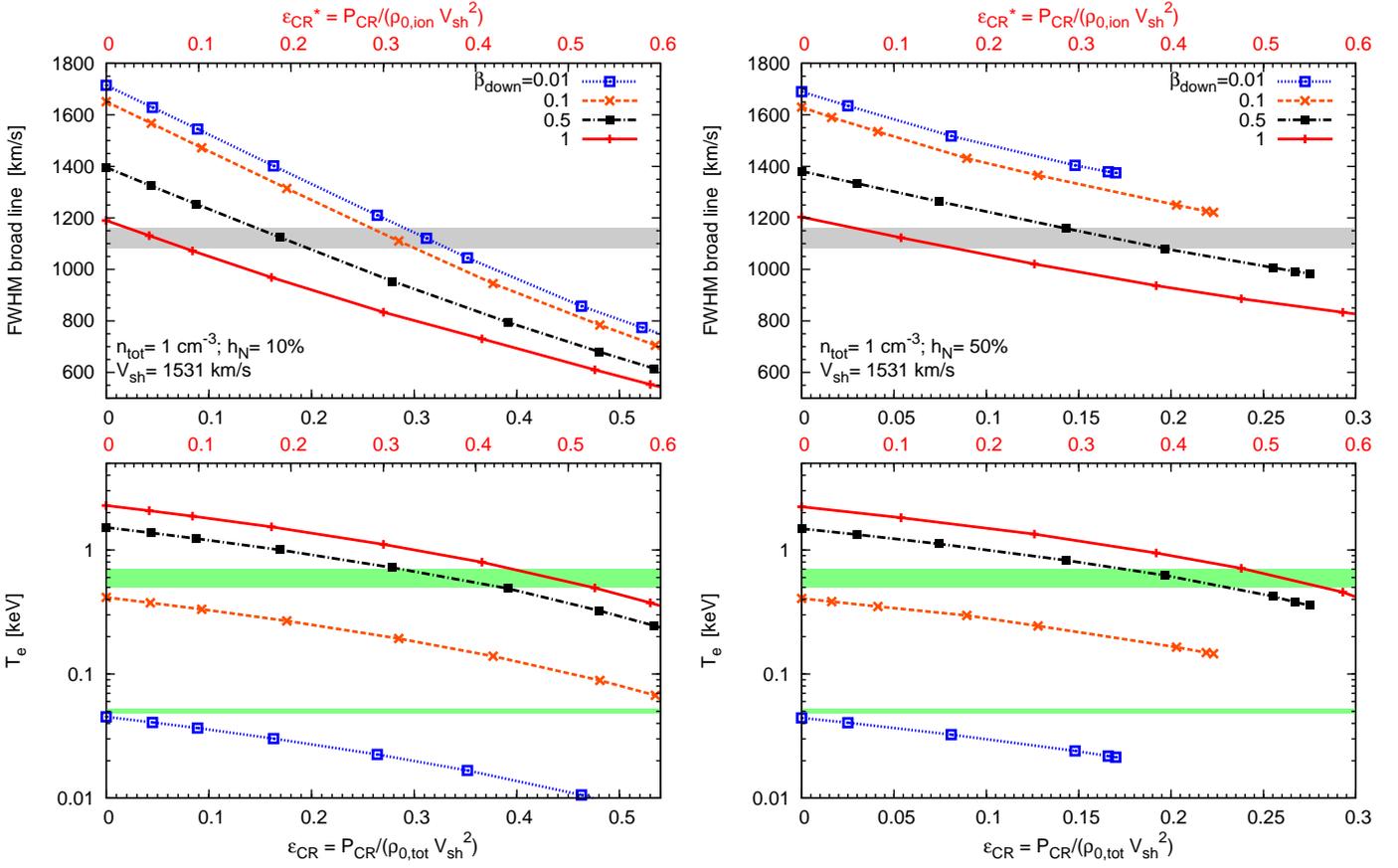}}
  \caption{{\it Top panels:} FWHM of the broad Balmer line as a function of the CR acceleration efficiency for a neutral fraction $h_{N}=10\%$ (left) and $h_{N}=50\%$ (right). Different lines show a different electron-ion equilibration level as labeled. The shaded area shows the FWHM as measured in the \SEout\ region by {\protect \cite{Helder13}} at 1 $\sigma$ level. {\it Bottom panels:} electron temperature immediately behind the shock as a function of the CR acceleration efficiency for $h_{N}=10\%$ (left) and $h_{N}=50\%$ (right). The shaded area shows $\Te$ as measured by {\protect \cite{Helder11}}, while the straight line at $\sim 0.05$ keV is the value estimated by the same authors for $\Te$ immediately behind the shock. All results refer to a shock velocity of $\Vsh= 1531\U{\kms}$ and a total gas density of $n_0= 1\U{\cmc}$.}
  \label{fig:FWHM_broad_CR}
\end{center}
\end{figure*}

In the upper panels of Fig. \ref{fig:FWHM_broad_CR} we show the FWHM as a function of the CR acceleration efficiency for different values of $\betadown$ and with $\hN=10\%$ (left panel) and $\hN=50\%$ (right panel), respectively. The calculations are fitted to the \SEout\ region (see again Fig.~1 in \cite{Helder11}; corresponding to Filament \#1 in Table 2 of \cite{Helder13}), which is most promising in terms of finding evidence for efficient particle acceleration. The shaded area represents the FWHM as measured by \cite{Helder11} in the same region, equal to $1120\pm 40\U{\kms}$. The optical proper motion for this filament, as measured by \cite{Helder13}, is $0.4\pm0.04$ arcsec in a period of 1135 days. For a source distance of 2.5 kpc, this corresponds to a shock velocity $1531\pm144 \kms$. 
Note that the CR acceleration efficiency in Fig.  \ref{fig:FWHM_broad_CR} is defined as $\epsCR=\PCR/\rhoztot\Vsh^{2}$, although a probably more meaningful definition from the physical point of view is $\epsCR^{*}=\PCR/\rhozion\Vsh^{2}$, where $\rhozion= (1-\hN)\,\rhoztot$ and $\hN$ are, respectively, the density of ionized material and the neutral fraction at upstream infinity. One should keep in mind that collisionless shocks only act on the ionized component of the background fluid and that only ions can take part in the acceleration process. The two bottom panels of Fig. \ref{fig:FWHM_broad_CR} show instead the electron temperature deduced for $\hN=10\%$ (left) and $\hN=50\%$ (right) respectively, compared with (shaded area) the electron temperature as measured by \cite{Helder11}. This measurement of the electron temperature is based on observations of the X-ray emission in the region of the \SEout\ Balmer filament, but it refers to a somewhat larger region that where Balmer emission is measured. \cite{Helder11} argued that the actual electron temperature behind the shock might be appreciably lower and tried to estimate this temperature by accounting for the effect of Coulomb scattering between electrons and ions. Clearly, this correction is strongly model dependent in that one needs to know the history of the shock evolution (expansion in an underdense bubble may reduce this effect, for instance). The estimate of the electron temperature immediately behind the shock as derived by \cite{Helder11} is shown in the two bottom panels of Fig. \ref{fig:FWHM_broad_CR} as a horizontal line at $\sim 0.05\U{keV}$. The region between the shaded area and the horizontal line can be considered as a reasonable estimate of the actual postshock electron temperature in the \SEout\ region of RCW86. On the other hand, given the uncertainties in the shock dynamics, we treated the measured electron temperature (shaded area) as an upper limit and used the horizontal line only for {\it a posteriori} checks.

We first discuss the case of lower neutral fraction, $\hN=10\%$ (left panels in Fig. \ref{fig:FWHM_broad_CR}). In the following we refer to this case as {\it Case A}. From the top panel we see that CR acceleration efficiencies between $\sim 5\%$ and $\sim 35\%$ are allowed, depending on the level of electron-ion equilibration, with higher efficiencies needed for lower values of $\betadown$. A value $\epsCR\sim 5\%$ is obtained for full equilibration, $\betadown=1$. However, the lower left panel in Fig. \ref{fig:FWHM_broad_CR} clearly shows that such a low level of CR acceleration would imply a too high electron temperature, incompatible with the value measured by \cite{Helder11}. Similar considerations hold for $\betadown=0.5$. For low values of $\betadown$, namely when electrons are much colder than ions behind the shock, the top panel shows that CR acceleration efficiencies on the order of $\sim 30\%$ are required, which appear to be compatible with the upper limit on the postshock electron temperature shown in the lower panel. We note that for $\betadown\leq 0.01$ the inferred electron temperature is even lower than the value $0.05$ keV inferred by \cite{Helder11} after correcting for the effect of Coulomb scattering. However, as discussed above, this value is very model dependent and should be taken with much caution. 

The case with a higher neutral fraction, $\hN=50\%$, is illustrated in the right panels of Fig. \ref{fig:FWHM_broad_CR} and shows some new interesting aspects. We refer to this case as {\it Case B}. For reasons that will be made clear below, some of the curves illustrating our results appear to be truncated at some level of CR acceleration efficiency. This implies that for sufficiently low values of $\betadown$ there is no intersection between the curves and the shadowed region showing the measured FWHM, so that these values of $\betadown$ are not compatible with the measured width. In general, for $\hN=50\%$, one can have CR acceleration with $\epsCR\sim 5$\,--\,$15\%$ only for $\betadown\sim 0.5$\,--\,1. These cases are also compatible with the upper limit on the electron temperature, as shown in the bottom right panel of Fig. \ref{fig:FWHM_broad_CR}.

The physical reason for this sort of saturation of the CR acceleration efficiency is very interesting and is related to the neutral return flux, described by \cite{paperI}. Note that for a distance to SNR RCW~86 of 2.5~kpc, the proper motion of the \SEout\ region is best fit with a shock velocity of $\sim 1530\U{\kms}$, sufficiently low that the cross-section for CE is large and the phenomenon of neutral return flux becomes of the utmost importance: a cold, fast neutral atom from upstream, penetrating the downstream region and suffering a CE reaction with a hot ion there, has a finite probability to give rise to a neutral moving upstream and release there its momentum and energy. \cite{paperI} discussed the implications of this phenomenon: the upstream ion plasma is slowed down and heated by the neutral return flux, thereby reducing the Mach number and weakening the shock. \cite{paperI} also calculated the modification of the spectrum of test particles accelerated at such shocks, finding that for a low shock speed the spectrum may become very steep at low energies. When the spectrum becomes softer than $p^{-5}$, the energetics of accelerated particles is dominated by particles with momentum close to the injection momentum. At this point, it becomes very hard to channel more energy into accelerated particles without smoothing the shock, hence the solution of the problem cannot be found. To be precise, this implies that a stationary solution cannot be found. It is possible that in nature this may induce some sort of intermittency in the shock structure. As expected, the effects of the neutral induced flux are more evident for larger neutral fractions, which is the reason why this phenomenon is visible only in the right panels of Fig. \ref{fig:FWHM_broad_CR}.

\begin{figure}
\begin{center}
{\includegraphics[width=\sizeFig\linewidth]{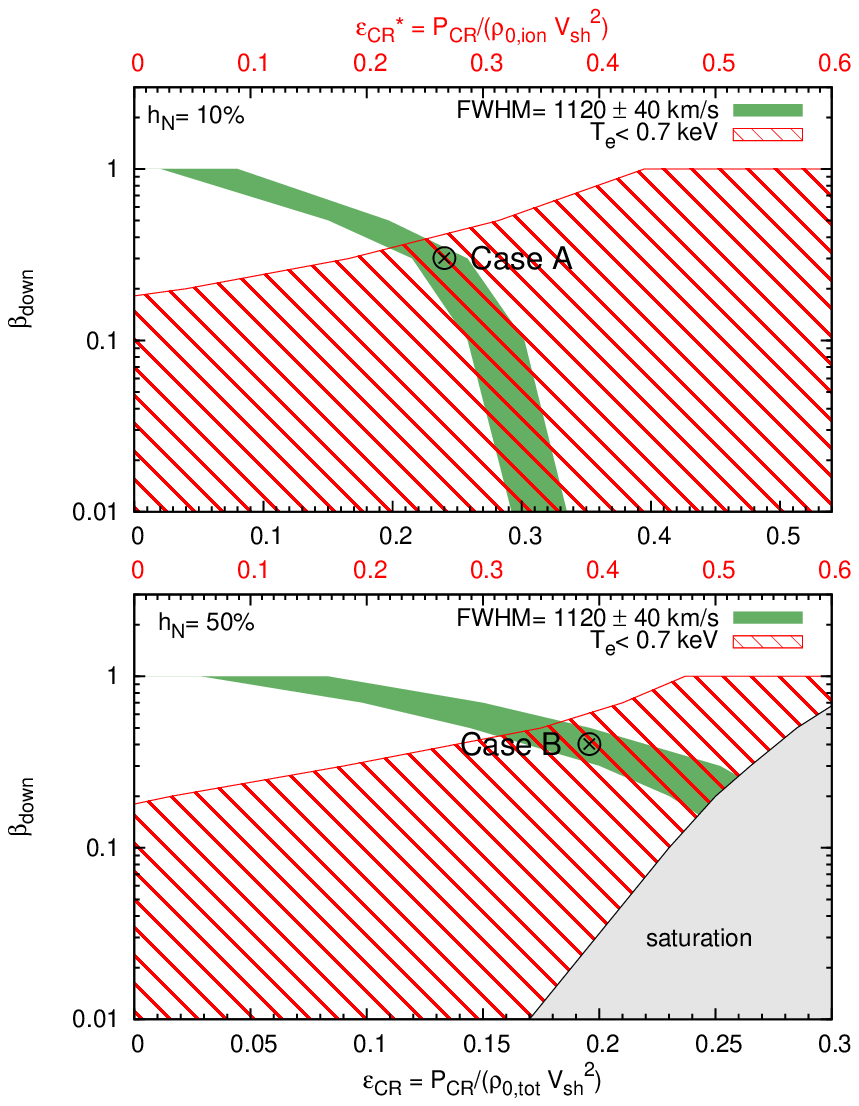}}
  \caption{Allowed regions of the parameter space $\beta_{\rm down}$-$\epsilon_{\rm CR}$ using different constraints. The allowed parameter space is given by the intersection of the green shaded area (constrained by the measured FWHM) and the hatched region (constrained by the upper limit on the electron temperature). The top and bottom panels refer to $h_{N}=10\%$ and $h_{N}=50\%$, respectively. The values of the parameters are the same as in Fig.~\ref{fig:FWHM_broad_CR}. In the lower panel the dashed gray region shows the saturation region. The crosses labeled  Case A and Case B refer to the cases analyzed in Figs.~\ref{fig:spectra} and \ref{fig:Balmer_profile} (see text for explanation).}
  \label{fig:FWHM_broad_CR-fN01}
\end{center}
\end{figure}

This rather complex situation is well illustrated in Fig. \ref{fig:FWHM_broad_CR-fN01}, where the information deriving from the measurement of the FWHM and the electron temperature are combined to define the allowed parameter space in terms of $\betadown$ and CR acceleration efficiency. The upper (lower) panel refers to Case A (Case B). The shadowed region is the allowed parameter space based on our calculations and on the value of the FWHM measured by \cite{Helder11}. The hatched area shows the constraints on the plane $\epsCR$\,--\,$\betadown$ obtained by imposing that $T_{e}\leq 0.7\U{keV}$ (upper edge of the shadowed region in the bottom panels of Fig. \ref{fig:FWHM_broad_CR}) and $\betadown\leq 1$. The allowed regions of the parameter space are those resulting from the overlap of the two contours. In the lower panel (Case B) the gray region shows the values of $\epsCR$\,--\,$\betadown$ where a stationary solution cannot be found because of the effect of the neutral return flux. For Case A (lower neutral fraction) the whole allowed region corresponds to CR acceleration efficiency $\epsCR\sim 20$\,--\,$30\%$ and $\betadown\lesssim 0.3$. For Case B (higher neutral fraction), the allowed region corresponds to $\epsCR\sim 15$\,--\,$25\%$ and $\betadown\sim 0.1$\,--\,$0.4$. In both cases it appears that assuming effective CR acceleration allows a better fit to the combined data on FWHM and electron temperature. 

\begin{figure}
\begin{center}
{\includegraphics[width=\sizeFig\linewidth]{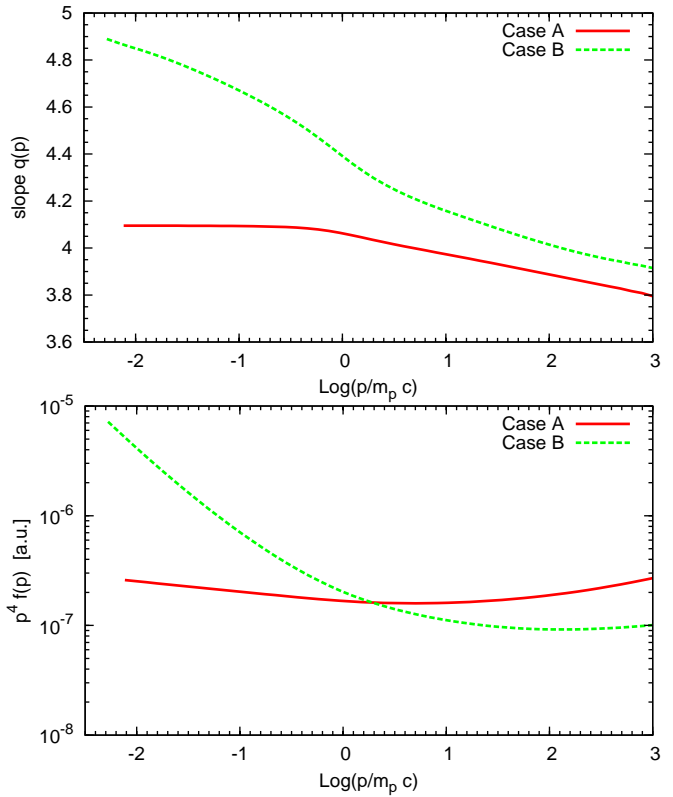}}
  \caption{Slope of the spectrum of accelerated particles (top panel) and CR spectrum multiplied by $p^4$ (lower panel) for Case A (solid lines) and Case B (dashed lines).}
  \label{fig:spectra}
\end{center}
\end{figure}

Our interpretation that the saturation effect discussed above is due to the neutral return flux is confirmed by the shape of the spectrum of the accelerated particles, illustrated in Fig. \ref{fig:spectra}: the upper panel shows the slope of the spectrum $q(p)=-\frac{d\ln f(p)}{d\ln p}$ as a function of momentum for Cases A and B, as labeled (here $f(p)$ is the distribution function of accelerated particles). The lower panel shows the actual spectrum of accelerated particles multiplied by $p^{4}$ (the test particle result would be a horizontal line in this plot). The spectra for Case A and Case B are calculated in the points shown by a cross in the allowed region of Fig. \ref{fig:FWHM_broad_CR-fN01}.

In Case A, when the neutral fraction is small, the spectrum is close to $f(p)\sim p^{-4}$, and the concavity (lower plot) is due to the CR induced precursor, expected for $\epsilon_{\rm CR}\sim 25\%$. On the other hand, in Case B, when the neutral fraction is higher, the slope of the spectrum at low particle momenta becomes close to 5, reflecting the strong effect of the neutral return flux. For particles with higher momentum, the diffusion length increases and the precursor region is increasingly more dominated by the CR reaction instead of by the neutral return flux, which spans distances from the shock on the order of a few CE interaction lengths. As a consequence, the spectrum of accelerated particles at higher momenta becomes even harder than $p^{-4}$. 
Moving rightward in the lower panel of Fig. \ref{fig:FWHM_broad_CR-fN01} within the allowed region, the low-energy spectrum of accelerated particles becomes increasingly steeper, and when the slope becomes $\gtrsim 5$, a stationary solution becomes impossible to find. 

\begin{figure}
\begin{center}
{\includegraphics[width=\sizeFig\linewidth]{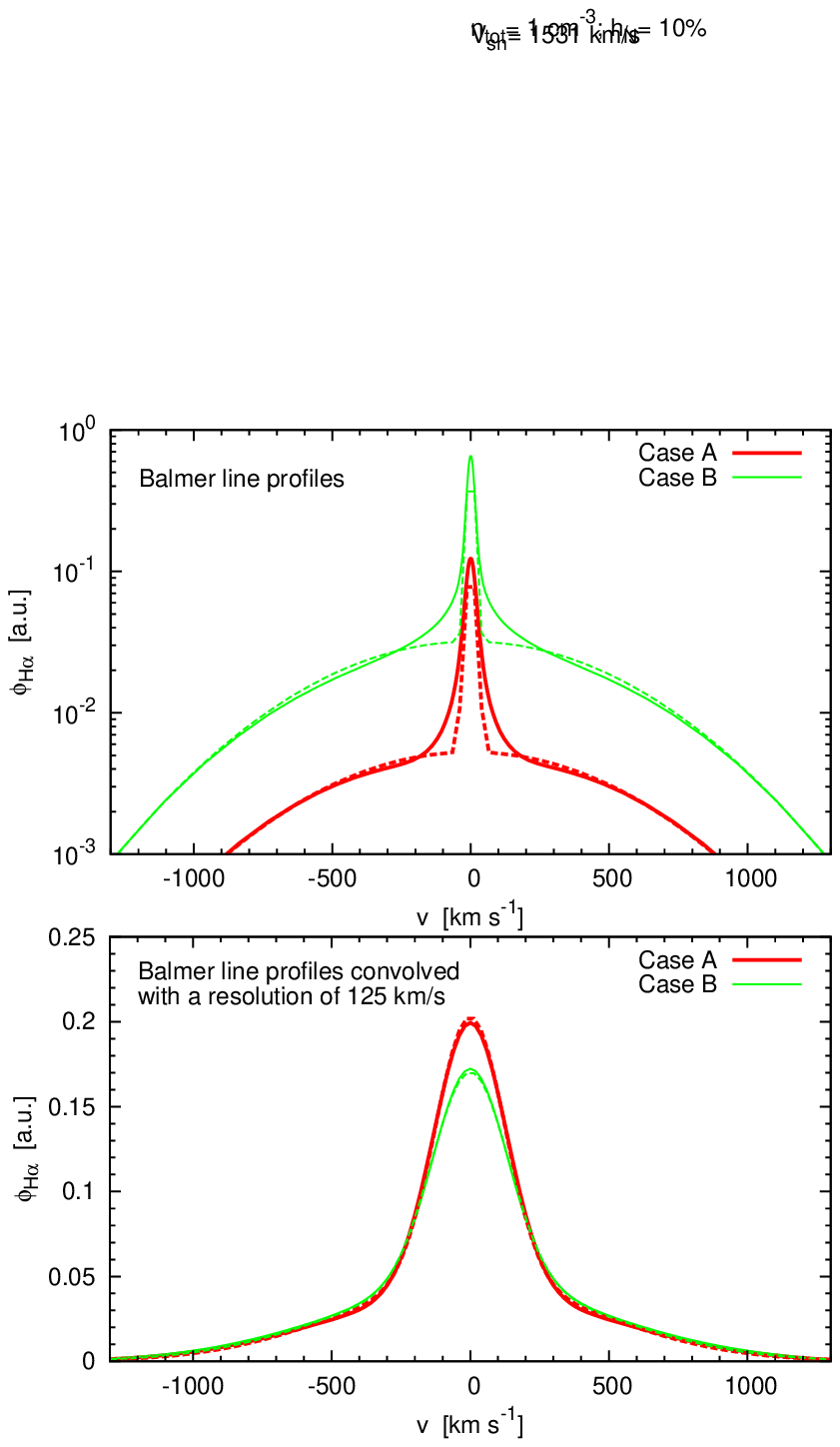}}
  \caption{Spatially integrated Balmer line emission for Case A (thick line) and Case B (thin line). The dashed lines show the best fit obtained using a two-Gaussian fit. In the bottom panel the Balmer emission is convolved with an instrumental resolution of $125\U{\kms}$, which corresponds to a FWHM of $350\U{\kms}$. In this case the wings disappear and the emission can be fitted well with two-Gaussian curves (dashed lines).}
  \label{fig:Balmer_profile}
\end{center}
\end{figure}

As discussed by \cite{paperI} and \cite{paperII}, the neutral return flux leads to the generation of a population of neutrals upstream of the shock with a temperature that is intermediate between the cold neutrals in the ISM (upstream infinity) and the hot neutrals present downstream due to CE with hot ions. This reflects in a modification of the profile of the Balmer line, which in addition to the well-known broad and narrow components develops an intermediate component, which typically has a FWHM of a few hundred km s$^{-1}$. 
In Fig.\ref{fig:Balmer_profile} we show the spatially integrated Balmer emission from RCW~86 for Cases A and B and for the values of $\beta_{\rm down}$ and $\epsilon_{\rm CR}$ indicated by the crosses in the allowed regions of Fig. \ref{fig:FWHM_broad_CR-fN01}. In the top panel the calculated emission is plotted together with a fit obtained using two Gaussians (dashed lines). It is easy to see that the quality of the fit is very poor, demonstrating that an additional component with intermediate width is needed. In the lower panel of Fig. \ref{fig:Balmer_profile} we plot the Balmer-line emission (as in the upper panel) convolved with a typical velocity resolution of $125\U{\kms}$. This illustrates how this type of observation would hide the presence of the intermediate Balmer line (having a width of $\sim 350\U{\kms}$), and in fact the total emission after convolution can easily be fitted with two Gaussians without an additional component. This fact clearly indicates that from the observational point of view it is essential to have measurements with a high-velocity resolution to identify the brightness and width of the intermediate Balmer line, which bears a signature of the neutral return flux. 

\section{Conclusions}
\label{sec:conclusion}

The quest for the origin of Galactic CRs remains open. The search for their origin is difficult and all new observational input need to be considered very carefully. Based mainly on the detection of gamma rays from several SNRs and on the morphology of the X-ray emission, which suggests strong magnetic field amplification, SNRs are considered, now more than ever before, as the most likely source of the bulk of CRs. Because the shocks produced by SN explosions often propagate in partially ionized gas, the emission lines of neutral hydrogen have recently been recognized as a possible diagnostic tool for CRs. This possibility has been put forward based on the fact that a shock that is accelerating particles is expected to be less effective in heating the background plasma, which is reflected in a Balmer-line emission with a smaller width than in the absence of CR acceleration. 

Here we used the theory of collisionless shocks in the presence of neutral hydrogen \cite[]{paperI,paperIII} to calculate the structure of the shock in SNR RCW~86 and infer the efficiency of CR acceleration in this remnant. The same formalism had previously been applied to SNR 0509\,--\,67.5 in the large Magellanic Cloud \cite[]{paperIV}. 

The recent measurements of proper motion carried out by \cite{Helder13} in several portions of the shock in RCW~86 have made it possible to achieve a quantitative estimate of the CR acceleration efficiency, although the uncertainty induced by the unknown distance to this SNR is very large. We found that the results of our calculations of the FWHM of the broad Balmer line are compatible with the absence of CR acceleration if the source distance is 2~kpc. On the other hand, for a source distance of 3~kpc, all measurements of the FWHM in \cite{Helder13} seem to require efficient CR acceleration. 

We assumed a fiducial distance to the source of 2.5 kpc and we restricted our calculations to the shock sector labeled \SEout\ in the paper by \cite{Helder11}, where the quality of the data is highest, showing evidence for particle acceleration. The measured proper motion for this region led to an estimated shock velocity of $\sim 1500\U{\kms}$. For a neutral fraction of 10\% we found that CR acceleration efficiencies between 5\% and 30\% are required to fit the measured FWHM, where the uncertainty is due to the unknown level of electron-ion equilibration behind the shock. \cite{Helder11} carried out X-ray measurements that allowed them to infer an upper limit to the postshock electron temperature of $T_{e}\leq 0.7\U{keV}$. We used this information to additionally constrain the parameter space: we found that to fit the data one has to require $\betadown\leq 0.3$ and CR acceleration efficiency 20\,--\,30\%. 

For a distance of 2.5~kpc and a neutral fraction of 50\% one is forced to assume $\betadown\sim 0.3$\,--\,0.5 and CR acceleration efficiency $\sim 15$\,--\,20\%. Interestingly, a stationary solution to the problem cannot be found outside this region. The physical reason for this phenomenon is that for a high neutral fraction the neutral return flux discussed by \cite{paperI} strongly affects the shock structure: the energy and momentum deposited by neutrals upstream of the shock are such that the shock is weakened and the spectrum of accelerated particles becomes very steep, to the point that the non-thermal energy is dominated by particles with a momentum close to the injection momentum. In these conditions it is very hard to increase the CR energy density any more and the problem does no longer have a stationary solution. 

This phenomenon is very important, since it suggests that for low shock velocities the FWHM of the broad Balmer line can be appreciably reduced (thereby suggesting effective CR acceleration) with no necessary implication for the non-thermal emission from the shock region: the spectrum of accelerated particles at the shock is so steep that one should not expect prominent gamma-ray emission or non-thermal X-ray emission from this region. However, this conclusion applies to cases of high neutral fraction, which are probably somewhat extreme. For lower neutral fraction, more appropriate to the ISM, the spectra are close to $p^{-4}$ and the inferred CR acceleration efficiency does reflect the actual CR content of the region where the FWHM of the Balmer line is measured. 

An independent test of the effect of the neutral return flux can be carried out by measuring the detailed structure of the Balmer line: the neutral return flux results in the production of a component of the Balmer line \cite[]{paperII}, which is intermediate between the narrow and the broad line, with a width of $\sim 350$ km s$^{-1}$. Such an additional component is easily hidden within the narrow or the broad component, depending on the velocity resolution adopted for the measurement.

\begin{acknowledgements}
We are grateful to D. Caprioli for discussions on the topic. This work was partially funded through grant PRIN INAF 2010, PRIN INAF 2012 and ASTRI. 
\end{acknowledgements}


\begin{thebibliography}{99}


\bibitem[\protect\citeauthoryear{Aharonian et al.}{2009}]{HESS09}
  Aharonian, F. et al. (HESS collaboration) 2009, ApJ, 692, 1500

\bibitem[\protect\citeauthoryear{Amato \& Blasi}{2005}]{AmatoBlasi05}
  Amato, E., \& Blasi, P., 2005, MNRAS, 364, L76

\bibitem[\protect\citeauthoryear{Amato \& Blasi}{2006}]{AmatoBlasi06}
  Amato, E., \& Blasi, P., 2006, \mnras, 371, 1251

\bibitem[\protect\citeauthoryear{Badenes et al.}{2007}]{Badenes07}
  Badenes, C., Hughes, J.~P., Bravo, E., Langer, N. 2007, ApJ, 662, 472
  
\bibitem[\protect\citeauthoryear{Bamba et al.}{2000}]{Bamba00}
  Bamba, A., Koyama, K., \& Hiroshi, T. 2000, PASJ, 52, 1157

\bibitem[\protect\citeauthoryear{Blasi et al.}{2012}]{paperI}
  Blasi, P., Morlino G., Bandiera R., Amato, E. \& Caprioli, D. 2012, ApJ, 755, 121 [Paper I]

\bibitem[\protect\citeauthoryear{Borkowski et al.}{2001}]{Borkowski01}
  Borkowski K. J., Rho, J., Reynolds, S. P., Dyer, K. K. 2001, ApJ, 550, 334
 
\bibitem[\protect\citeauthoryear{Caprioli et al.}{2008}]{Caprioli08}
  Caprioli, D., Blasi, P., Amato, E., \& Vietri, M. 2008, ApJ, 679, L139

\bibitem[\protect\citeauthoryear{Chevalier \& Raymond}{1978}]{Chevalier78}
 Chevalier, R. A., \& Raymond, J. C. 1978, ApJ, 225, L27

\bibitem[\protect\citeauthoryear{Claas et al.}{1989}]{Claas89}
 Claas, J. J., Kaastra, J. S., Smith, A., Peacock, A. \& de Korte, P. A. J., 1989, ApJ, 337, 399

\bibitem[\protect\citeauthoryear{Dickel, Strom \& Milne}{2001}]{Dickel01}
 Dickel, J. R., Strom, R. G., \& Milne, D. K., 2001, ApJ, 546, 447

\bibitem[\protect\citeauthoryear{Ghavamian et al.} {2013}]{2013SSRv.tmp.75G}
 Ghavamian P., Schwartz S.J., Mitchell J., Masters A., Laming J.M., 2013
  {\it Electron-Ion Temperature Equilibration in Collisionless Shocks: The
  Supernova Remnant-Solar Wind Connection}, Preprint ArXive:1305.6617

\bibitem[\protect\citeauthoryear{Greidanus \& Strom}{1990}]{Greidanus90}
 Greidanus, H., \& Strom, R. G., 1990, A\&A, 240, 385

\bibitem[\protect\citeauthoryear{Gvaramadze \& Vikhlinin}{2003}]{Gvaramadze03}
 Gvaramadze, V.~V. \& Vikhlinin, A.~A., 2003, A\&A, 401, 625
	
\bibitem[\protect\citeauthoryear{Helder et al.}{2009}]{Helder09} 
 Helder, E. A., Vink, J., Bassa, C. G., Bamba, A., Bleeker, J. A. M., Funk, S., Ghavamian, P., van
 der Heyden, K. J., Verbunt, F., Yamazaki, R., 2009, Science, 325, 719
 
\bibitem[\protect\citeauthoryear{Helder et al.}{2011}]{Helder11}
  Helder, E. A., Vink, J., \& Bassa, C. G. 2011, ApJ, 737, 85 
  
\bibitem[\protect\citeauthoryear{Helder et al.}{2013}]{Helder13}
  Helder, E. A., Vink, J., Bamba, A., Bleeker, J. A. M., Burrows, D. N., Ghavamian, P. and Yamazaki, R. 2013, MNRAS, 435, 910

\bibitem[\protect\citeauthoryear{ Leibowitz \& Danziger}{1983}]{Leibowitz83}
  Leibowitz, E. M. \& Danziger, I. J. 1983, MNRAS, 204, 273

\bibitem[\protect\citeauthoryear{Malkov \& Drury}{2001}]{maldrury}
 Malkov, M.A. \& Drury, L. O'C., 2001, Rep.~Prog.~Phys., 64, 429

\bibitem[\protect\citeauthoryear{Morlino et al.}{2012}]{paperII}
  Morlino, G., Bandiera, R., Blasi, P. \& Amato, E. 2012, ApJ, 760, 137 [Paper II]

\bibitem[\protect\citeauthoryear{Morlino et al.}{2013a}]{paperIII}
  Morlino G., Blasi, P., Bandiera R., \& Amato, E. 2013, ApJ, 768, 148 [Paper III]

\bibitem[\protect\citeauthoryear{Morlino et al.}{2013b}]{paperIV}
  Morlino G., Blasi, P., Bandiera R., \& Amato, E. 2013, A\&A, 557, 142 

\bibitem[\protect\citeauthoryear{Petruk}{1999}]{Petruk99}
  Petruk, O. 1999, A\&A, 346, 961

\bibitem[\protect\citeauthoryear{Raymond et al.}{2011}]{Raymond11}
  Raymond, J. C., Vink, J., Helder, E. A, \& de Laat, A., 2011, ApJ, 731, L14

\bibitem[\protect\citeauthoryear{Rosado et al.}{1996}]{Rosado96}
  Rosado, M., Ambrocio-Cruz, P., Le Coarer, E., Marcelin, M.. 1996, A\&A, 315, 243

\bibitem[\protect\citeauthoryear{Smith}{1997}]{Smith97}  
  Smith, R.C. 1997, AJ, 114, 2664

\bibitem[\protect\citeauthoryear{van Adelsberg et al.}{2008}]{vanAdel08} 
 van Adelsberg, M., Heng, K., McCray, R., \& Raymond, J. C. 2008, \apj, 689, 1089

\bibitem[\protect\citeauthoryear{Vink et al.}{2006}]{Vink06}
  Vink, J., Bleeker, J., van der Heyden, K., Bykov, A., \& Yamakazi, R. 2006, ApJ, 648, 33

\bibitem[\protect\citeauthoryear{Vink et al.}{1997}]{Vink97}
  Vink, J., Kaastra, J. S., \& Bleeker, J. A. M. 1997, A\&A, 328, 628

\bibitem[\protect\citeauthoryear{Wagner et al.}{2009}]{Wagner09}
  Wagner, A. Y., Lee, J.-J., Raymond, J. C., Hartquist, T. W., \& Falle, S. A. E. G. 2009, ApJ, 690, 1412

\bibitem[\protect\citeauthoryear{Westerlung}{1969}]{Westerlung69}
  Westerlung, B. E. 1969, AJ, 74, 879

\bibitem[\protect\citeauthoryear{Williams et al.}{2011}]{Williams11}
  Williams, B. J., Blair, W. P., Blondin, J. M. et al. 2011, ApJ, 741, 96

\end{thebibliography}
\end{document}